\documentclass{article}
\usepackage{spconf,amsmath,graphicx, multirow,amsfonts}
\usepackage{amssymb,verbatim,xcolor,bm,svg}
\usepackage{algorithm,setspace, enumitem}
\usepackage{caption}
\usepackage{subcaption}
\usepackage{hyperref}
\usepackage[noend]{algpseudocode}
\usepackage[normalem]{ulem}
\input{def.set}

\title{Greedy Blockwise Residual Learning\\for Scalable Monaural Source Separation}
\title{BLOOM-Net: Blockwise Optimization for Masking Networks\\Toward Scalable and Efficient Speech Enhancement}

\name{Sunwoo Kim, Minje Kim\thanks{This material is based upon work supported in part by the National Science Foundation under Grant Numbers 1909509 and 2046963.}}
\address{Indiana University, Department of Intelligent Systems Engineering, Bloomington, IN, USA 47408}
\begin{document}
\ninept
\maketitle
\begin{abstract}
In this paper, we present a blockwise optimization method for masking-based networks (BLOOM-Net) for training scalable speech enhancement networks. Here, we design our network with a residual learning scheme and train the internal separator blocks sequentially to obtain a scalable masking-based deep neural network for speech enhancement. 
Its scalability lets it dynamically adjust the run-time complexity depending on the test time environment. To this end, we modularize our models in that they can flexibly accommodate varying needs for enhancement performance and constraints on the resources,
incurring minimal memory or training overhead due to the added scalability. Our experiments on speech enhancement demonstrate that the proposed blockwise optimization method achieves the desired scalability with only a slight performance degradation compared to corresponding models trained end-to-end. 
\end{abstract}
\section{Introduction}
\label{sec:intro}








Deep learning-based supervised methods have dramatically boosted single-channel source separation performances in recent years. Typically, effective and dominating deep learning solutions operate by estimating masks, 
either on the time-frequency (TF) representations, such as the short-time Fourier transform (STFT) \cite{NarayananA2013icassp, WilliamsonD2016complex, HersheyJ2016icassp} 
or recent models that learn a \textit{separator} module that applies masks in the latent feature space. The latter models have improved the state of the art as the learned feature space allows them to bypass limits imposed by TF-domain solutions (e.g., time-frequency resolution trade-off, using noisy phase or dealing with phase estimation, etc.) in addition to the advanced separator module's architecture. 
Various architectures have been proposed with each progressive model showing relative improvements: fully-convolutional Conv-TasNet \cite{LuoY2019conv-tasnet} that initially popularized the time-domain approach, dual-path recurrent neural networks (DPRNN) that enabled long-term sequence modeling \cite{YiL2020dualpathRNN}, and Transformer and Conformer-based models that overcome limitations from convolutional neural networks (CNN) and RNN based approaches (e.g., limitations of receptive fields and extensive recurrent connections) \cite{Chen2020dual,Subakan2021attention,Chen2021conformer, Koizumi2021dfconformer}. 

However, a major drawback of aforementioned deep learning solutions is the complexity of models. Heavy memory occupancy and especially their exorbitant computational cost makes them impractical for deployment onto resource-constrained devices. 

Model compression methods offer effective solutions to this problem by reducing complexity of neural network architectures while minimizing their drop in generalization performance. There are various model compression methods such as quantizing model parameters using low-bit resolution fixed-point representations and/or pruning less important network components 
\cite{HanS2016iclr, KimSW2019icassp}
, simplifying convolutional operations \cite{HowardAG2017Mobilenets}, grouping RNN's intermediate tensor representations into smaller blocks \cite{Luo2021group}, multi-resolution features via successive downsampling and resampling \cite{TzinisE2020sudormrf}, distilling knowledge from a larger network to improve the performance of its corresponding compressed model  \cite{HintonG2015arxiv, KimSW2021waspaa}, etc. These compressed models are typically designed to minimize the inference complexity targeting the low-resource environment, thus not being able to scale up to challenging separation problems. Eventually, in order for a legacy system to be scalable, a range of versions need to be retained in a device, increasing the total spatial complexity.


We argue that a scalable and efficient system must cover a wide resource-related diversity in edge computing via an \textit{adaptive model architecture} rather than simply enumerating various model architectures. The scalable, thus adaptive systems can be commonly found in coding applications. In \cite{ZhenK2019interspeech}, the cross-module residual learning scheme enabled greedy module-wise neural codec learning, where a deep autoencoder is trained to model the residual signal that its preceding autoencoder fails to model. The system can preserve the order of relative importance of the participating autoencoder modules, that gives scalability to the system, i.e., the first part of the bitstream is more important than the rest. 
SoundStream audio codec features bitrate scalability, allowing the codec to adapt to the network conditions that can vary while transmitting signals, too \cite{Zeghidour2021soundstream}. However, these models are specifically for signal compression, seeking scalability within their resulting bitstreams, thus not suitable for other applications.
Meanwhile, the once-for-all (OFA) scheme provides a general-purpose adaptive training mechanism that learns multiple compressed variants of a model via a single training task \cite{Cai2019once}. However, it does not provide a single architecture that scales to different test environments freely as we propose. 

Likewise, we envision a scalable speech enhancement model that changes its operation modes, ranging from an energy-efficient version to a performance-boosted one. In doing so, instead of preparing each different model in the device, a scalable model provides a flexible structure that adjusts its performance and complexity per resource constraint.
To this end, we propose a scalable time-domain architecture for speech enhancement: BLOckwise Optimization for Masking networks (BLOOM-Net). It is a greedy residual learning strategy to train individual blocks sequentially. Since each block is trained to improve the previous block's speech enhancement result further, the deployment can optionally choose to use only the first few blocks depending on the available computational resources. Although it has been known that a sequence of two heterogeneous speech enhancement processes is effective \cite{ZhaoY2018two-stage, HaoX2020masking-inpainting}, our approach differs from that literature in that (a) ours focuses on the architectural innovation rather than a concatenation of two heterogeneous models (b) the proposed model eventually achieves the scalability in the feature space rather than in the raw signal domain.
BLOOM-Net shows competitive performance compared to their end-to-end counterparts, while providing the additional advantage that its modularized blocks are easy to attach or detach for scalability.





\begin{figure}
    \centering
    \begin{subfigure}[b]{0.495\columnwidth}
         \centering
         \includegraphics[width=\textwidth]{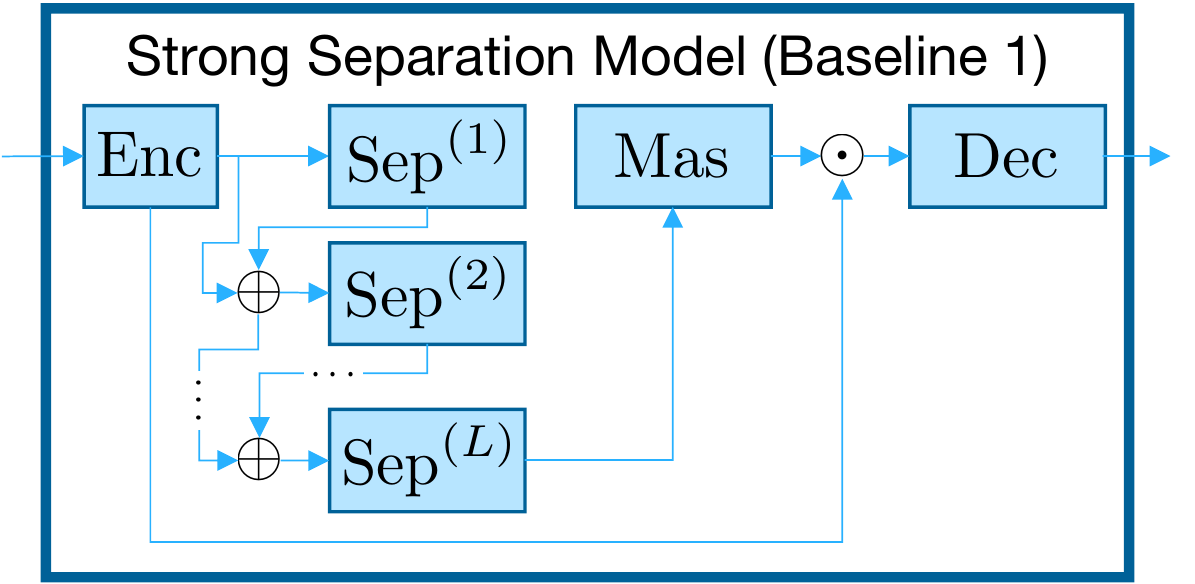}
         \caption{}
         \label{fig:strong_separation}
     \end{subfigure}
     \begin{subfigure}[b]{0.495\columnwidth}
         \centering
         \includegraphics[width=\textwidth]{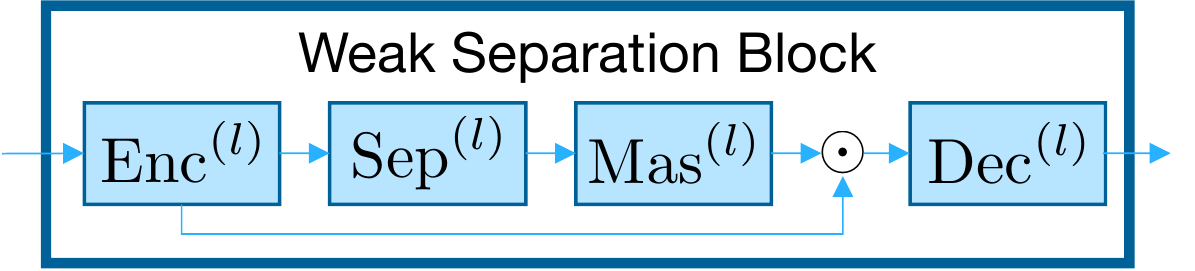}
         \caption{}
         \label{fig:weak_block}
     \end{subfigure}
     \caption{(a) The baseline separation model (b) a weak separation block}
\end{figure}

\section{Methodologies}
\label{sec:method}

\subsection{Baseline 1: The Time-Domain Separation Model}\label{sec:bs1}
Our baseline source separation model adopts the common structure found in time-domain source separation networks that consists of an encoder, separator, masker, and decoder as shown in Fig. \ref{fig:strong_separation}. 
First, the input $\bx\in\mathbb{R}^{T}$ with $T$ samples in the time domain is fed to
the encoder, which is a function that transforms the input into a latent representation: $\text{Enc}:\bx \rightarrow \bh\in\mathbb{R}^D$.  We posit that it is a small neural network module, e.g., a 1D convolutional layer followed by a nonlinear activation function.
Then, the latent representation is fed into the separator module: $\text{Sep}:\bh \rightarrow \bz\in\mathbb{R}^K$, whose output $\bz$ is used as input to the mask estimation module $\text{Mas}:\bz\rightarrow\boldm\in\Real^D$. 
Finally, the mask is applied to the encoded mixture representation $\bh$ to retrieve the source-specific estimate of the latent representation $\tilde{\bh}$ (e.g., for the speech source), which is then decoded back to the time-domain estimate of the target source, $\text{Dec}:\tilde\bh\rightarrow\hat\bs\in\Real^T$. 

In the time-domain separation models, it is common to employ a large separator module with repeating structures indexed by $l$. It is also popular to merge the input and output of each block as input to the next one, performing residual learning as proposed in ResNet \cite{HeK2016cvpr}. These blocks are learned altogether in the state-of-the-art models as well as in Baseline 1, limiting the models' scalability.

\begin{figure}
    \centering
    \begin{subfigure}[b]{\columnwidth}
         \centering
         \includegraphics[width=.9\textwidth]{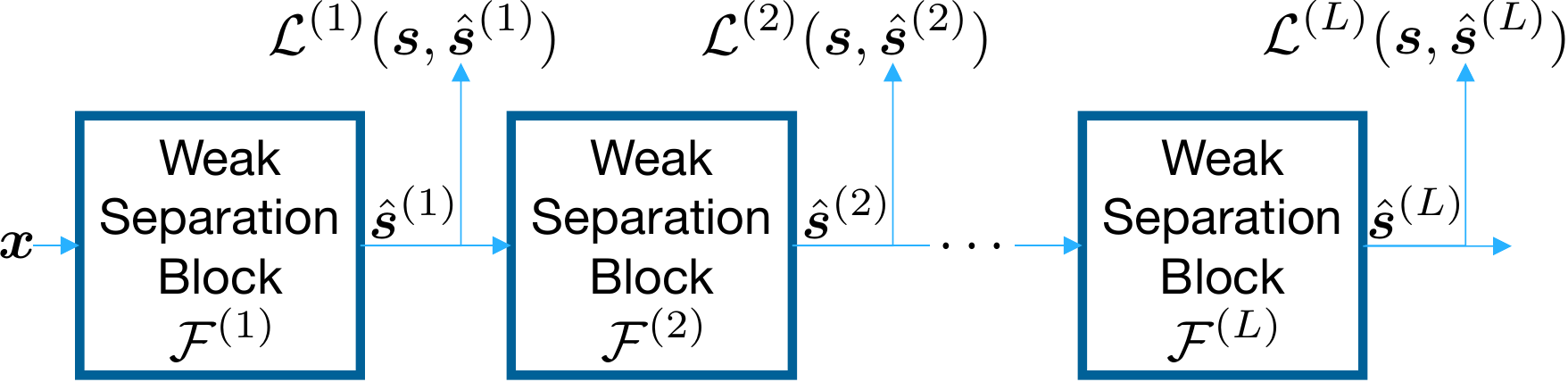}
         \caption{The time-domain blockwise optimization model (Baseline 2).}\vspace{0.1in}
         \label{fig:tdrl}
     \end{subfigure}
     \begin{subfigure}[b]{\columnwidth}
         \centering
         \includegraphics[width=\textwidth]{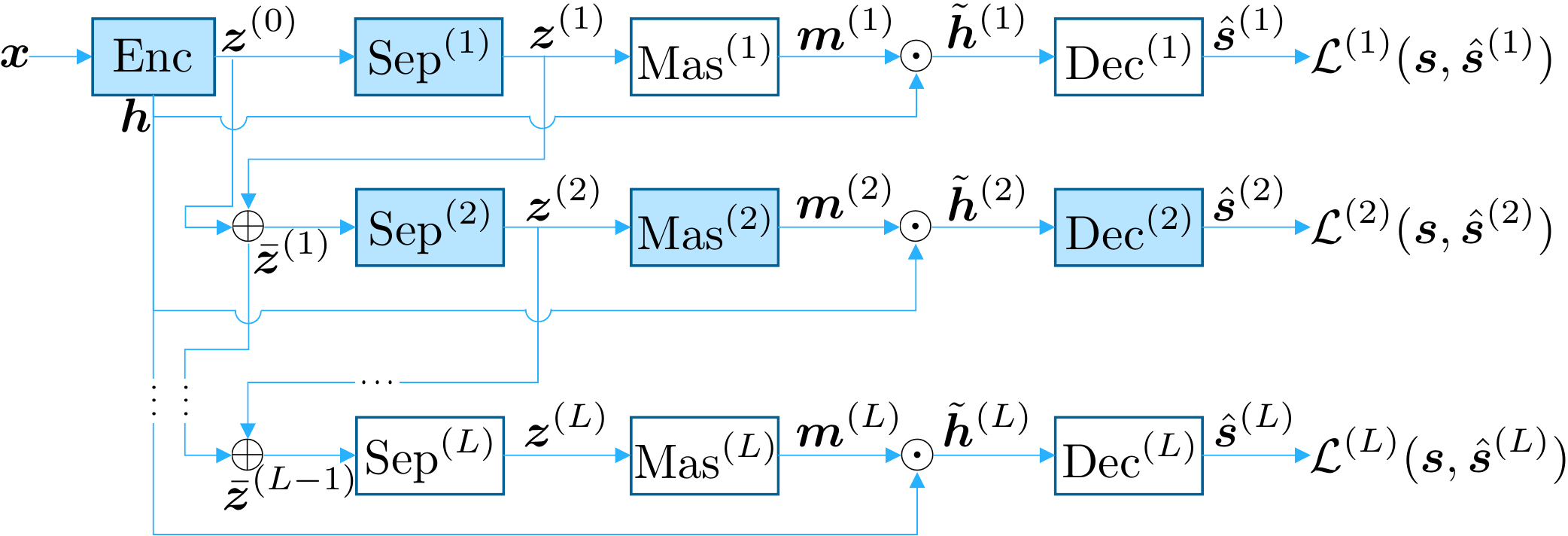}
         \caption{The proposed BLOOM-Net architecture.}
         \label{fig:fsrl}
     \end{subfigure}
     \caption{The scalable speech enhancement models. }
\end{figure}

\subsection{Baseline 2: Time-Domain Blockwise Optimization}\label{sec:bs2}

To build up our proposed BLOOM-Net model, we begin with a na\"ive concatenation approach as our second baseline. Fig. \ref{fig:weak_block} shows a \textit{weak} source separation block as a compromised version of Baseline 1 in Fig. \ref{fig:strong_separation}. Although it is with only one ResNet block for the separator module, it is still a legitimate stand-alone separation model. We assume that there are $L$ such weak separation blocks, each of which enhances its previous block's results as shown in Fig. \ref{fig:tdrl}. Let $\calF^{(l)}(\cdot)$ be the $l$-th weak separation block. It performs speech enhancement on $\hat\bs^{(l-1)}$, an output of the $(l-1)$-th block: $\calF^{(l)}:\hat\bs^{(l-1)}\rightarrow\hat\bs^{(l)}$. Note that the notion of a ``weak" separation block comes from the boosting methods that incrementally add weak learners \cite{FreundY1996adaboost}.

The second baseline provides a mechanism to serialize the speech enhancement models with an order of significance. If the $(l-1)$-th weak separation block leaves room for improvement, i.e., when the loss $\calL^{(l-1)}(\bs, \hat{\bs}^{(l-1)})$ is not sufficiently small, the next block $\calF^{(l)}(\hat\bs^{(l-1)})$ focuses on that sample and tries to improve. During the test time, suppose the device can afford only up to $\ell$ blocks. 
Since each of the $L$ blocks is sequentially trained with its own reconstruction loss $\calL^{(l)}(\bs, \hat{\bs}^{(l)})$, scalability is achieved by performing the inference on only the first $l \leq \ell$ weak separation blocks---the result is still a legitimate source estimate. 
In contrast, Baseline 1 would need 
to redundantly prepare all the scaled variants of the model in order to achieve the desired scalability. 

\subsection{BLOOM-Net: Blockwise Optimization in the Latent Space}\label{sec:bloom}
Baseline 2 exhibits a redundancy issue. During the test time, the chosen block should execute all four submodules to deliver a time-domain signal to its successive block. Instead, we propose a blockwise optimization scheme that works in the latent space, so that the serialization is done among the separator submodules. In that way, the system can avoid unnecessary masking, decoding, and encoding operations that repeat at every weak separation block in Baseline 2. 

Fig. \ref{fig:fsrl} describes the BLOOM-Net architecture, where $\text{Sep}^{(l)}$ performs residual learning in the sense that it relays the sum of its output $\bz^{(l)}$ and the input $\bar\bz^{(l-1)}$ to the next separator block:
\begin{equation}
    \bar\bz^{(l)} = \bz^{(l)} + \bar\bz^{(l-1)}, \text{ where } \bz^{(l)}=\text{Sep}^{(l)}\big(\bar\bz^{(l-1)}\big). 
\end{equation}
Note that the encoder's output $\bh=\bz^{(0)}$ is the input to $\text{Sep}^{(1)}$. We also induce the input to $\text{Sep}^{(L)}$ recursively: $\bar\bz^{(L-1)} = \sum_{l=0}^{L-1}\bz^{(l)}$.

BLOOM-Net performs block-specific masking and decoding just to compute the block-specific error, while the residual connections are defined among the latent variables $\bz^{(l)}$. The $l$-th masker and decoder works on the separator output $\bz^{(l)}$ as follows:
\begin{equation}
    \hat\bs^{(l)}\!=\!\text{Dec}^{(l)}\big(\tilde\bh^{(l)}\big),~\tilde\bh^{(l)}\!=\!\boldm^{(l)}\odot\bh^{(l)},~~\boldm^{(l)}\!=\!\text{Mas}^{(l)}\big(\bz^{(l)}\big). 
\end{equation}
The blockwise output $\hat\bs^{(l)}$ is then compared with the ground-truth source $\bs$ to compute the blockwise loss $\calL^{(l)}(\bs, \hat\bs^{(l)})$, ensuring the intermediate output is usable. 
Note that $\text{Enc}$ is shared and reused for all the sequence of blocks such that all blocks learn to estimate denoising masks on the same latent representation $\bh$. Meanwhile, $\text{Mas}^{(l)}$ and $\text{Dec}^{(l)}$ modules are block-specific and no longer updated once the $l$-th block is trained.

During the test time, 
the actual inference involves $\text{Enc}$, $\text{Sep}^{(l)}$ where $1\!\leq\!l\!\leq\!\ell$, $\text{Mas}^{(l)}$, and $\text{Dec}^{(l)}$, which are represented as shaded blocks in Fig. \ref{fig:fsrl}. Compared to Baseline 2, BLOOM-Net saves the cost of $\ell-1$ encoding, decoding, and masking operations. 
Although these blocks are lightweight, removing them improves not only the computational efficiency 
but also feature learning. 
The direct residual learning path in the latent feature space allows BLOOM-Net to learn a hierarchy of latent representations, while Baseline 2 is limited to concatenating shallow representations. 

Fine-tuning can further refine a fully trained BLOOM-Net. 
While BLOOM-Net provides the desired scalability via blockwise optimization, its greedy nature prevents the model from learning from the full picture, thus underperforming the theoretical upper bound. To this end, we propose to fine-tune BLOOM-Net, where all modules in all $L$ blocks are updated using the combination of all loss functions, i.e., $\sum_{l=1}^L\calL^{(l)}(\bs, \hat\bs^{(l)})$. This comprehensive fine-tuning significantly improves BLOOM-Net, making its performance near the theoretical upper bound.

\section{Experimental Setup}
\label{sec:setup}

\subsection{Datasets}


During training, we used clean speech samples from the Librispeech corpus \cite{PanayotovV2015Librispeech} and noise recordings from the MUSAN dataset \cite{SnyderD2015MUSAN}. We used \texttt{train-clean-100} and \texttt{dev-clean} subsets from Librispeech for training and validation respectively. We split MUSAN's \texttt{free-sound} subset at 80:20 ratio into training and validation partitions. 
For testing, we used unseen speech samples from Librispeech's \texttt{test-clean} and noise from MUSAN's \texttt{sound-bible}. 
Audio files are loaded at 16 kHz sampling rate and standardized to have a unit-variance. 
Noise samples are scaled to random input SNR levels uniformly chosen between -5 and 10 dB and added to speech signals to obtain noisy mixtures.

\subsection{Architecture and Training Details}
To investigate its application to time-domain audio separation networks \cite{LuoY2019conv-tasnet,YiL2020dualpathRNN,Chen2020dual,Subakan2021attention, LuoY2018tasnet}, our ResNet-based models are implemented as a simplified form of Conv-TasNet \cite{LuoY2019conv-tasnet}. 
We use the same encoder and decoder design as in \cite{LuoY2019conv-tasnet} with the same hyperparameters. 
Each separator module is defined as a residual block using 1-D convolutional layers similar to those in \cite{LuoY2019conv-tasnet}, but without dilation and intermediate skip-connections. 
Each residual block consists of a $1\times1$ convolution operation followed by a depthwise convolution operation, with parametric ReLU (PReLU) \cite{HeK2015cvpr} and global layer normalization (gLN) added after each convolution operation. 
We used the ConvTasNet implementation available in Asteroid \cite{ParienteM2020asteroid} and tuned the hyperparameters to construct our time-domain ResNet models.

We used the negative scale-invariant signal to distortion ratio (SI-SDR) as the loss function \cite{LeRouxJL2018sisdr}, defined as 
\begin{equation}
    \calL(\bs,\hat{\bs}) = -\text{SI-SDR}(\bs,\hat{\bs}) = -10\log_{10}\bigg(\frac{||\alpha \bs||^2}{||\alpha \bs - \hat{\bs}||^2}\bigg)
\end{equation}
where $\alpha = \frac{\hat{\bs}^\top \bs}{||\bs||^2}$ is a scaling factor. 
We train all models on 1-second long segments using a mini-batch size of 64. 
Adam is used as the optimizer with learning rate initialized to $1\times 10^{-4}$. 
Early stopping is applied and models with the lowest validation losses are used for final evaluation on the test mixtures. 

\subsection{Training Configurations}
The experiments are on two types of baselines and our proposed BLOOM-Net method. 
We additionally examine the impact of fine-tuning on BLOOM-Net. 
\begin{itemize}[noitemsep,topsep=0pt, leftmargin=0in, itemindent=.15in]
    \item \textit{Baseline 1 - Full}: The time-domain separation model (Sec. \ref{sec:bs1}) trained in a conventional end-to-end manner. It employs $L=6$ separator blocks from the beginning and learns all of them together. Although there are $L=6$ separators, removing one of them to reduce the complexity will cause a complete break down as the model is not scalable.
    \item \textit{Baseline 1 - Int.}: We also learn $L=6$ different versions of Baseline 1, each of which is an intermediate version,  containing up to $L=\{1,\ldots,6\}$ separators, respectively, e.g., when $L=3$ there are three separators that are trained altogether. Each of these models is the upper bound of its corresponding BLOOM-Net with a matching separator number. 
    \item \textit{Baseline 2}: The time-domain weak separation blocks (Sec. \ref{sec:bs2}). Individual blocks are trained as a stand-alone separation model one after another to promote scalability. 
    \item \textit{BLOOM}: BLOOM-Net in its basic setup (Sec. \ref{sec:bloom}). 
    \item \textit{BLOOM-FT}: BLOOM-Net, fine-tuned to minimize the combination of losses from all blocks. It overcomes the suboptimal performance caused by \textit{BLOOM}'s greedy training.
    \end{itemize}

\renewcommand{\arraystretch}{1.}
\begin{table}[t]
\centering
\caption{SI-SDR improvements of the competing models. 
Evaluations are with respect to the number of blocks $\ell$ chosen for inference.
}
\resizebox{\columnwidth}{!}{
\begin{tabular}{crrrrrr}
\hline
Method & $\ell=1$ & $\ell=2$ & $\ell=3$ & $\ell=4$ & $\ell=5$ & $\ell=6$\\
\hline\hline
\textit{Baseline 1 - Full} & -0.60 & -0.86 & -0.12 & 0.40 & 0.97 & 8.89\\
\hline
\textit{Baseline 1 - Int.} & 4.55 & 6.83 & 7.65 & 8.30 & 8.58 & 8.89\\
\hline
\textit{Baseline 2} & 4.55 & 5.01 & 5.17 & 5.25 & 5.25 & 5.25 \\
\hline
\textit{BLOOM} & 4.55 & 6.13 & 6.92 & 7.40 & 7.61 & 7.75 \\
\hline
\textit{BLOOM-FT} & 4.74 & 6.55 & 7.44 & 8.14 & 8.51 & 8.72 \\
\hline
\end{tabular}
\label{tab:perf}
}
\end{table}

\begin{figure*}[t]
        \centering
        \begin{subfigure}[t]{0.195\textwidth}
            \centering
            \includegraphics[height=1.220in]{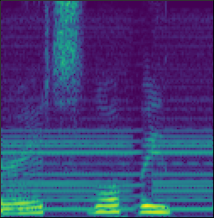}\vspace{0.03in}
            \caption[snr-5]%
            {{\small $\bx_1^{~}$ (SI-SDR: 7.63 dB)}}    
            \label{fig:snr-5}
        \end{subfigure}
        \hfill
        \begin{subfigure}[t]{0.195\textwidth}  
            \centering 
            \includegraphics[height=1.220in]{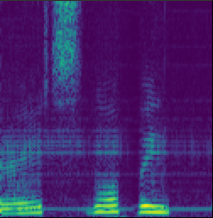}
            \caption[snr0]%
            {{\small $\hat{\bs}_1^{(2)}$ (SI-SDR: 12.12 dB)}}    
            \label{fig:snr0}
        \end{subfigure}
        \hfill
        \begin{subfigure}[t]{0.195\textwidth}   
            \centering 
            \includegraphics[height=1.220in]{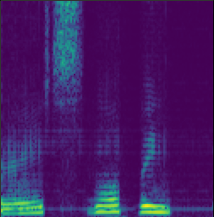}
            \caption[snr5]%
            {{\small $\hat{\bs}_1^{(4)}$ (SI-SDR: 13.13 dB)}}    
            \label{fig:snr5}
        \end{subfigure}
        \hfill
        \begin{subfigure}[t]{0.195\textwidth}   
            \centering 
            \includegraphics[height=1.220in]{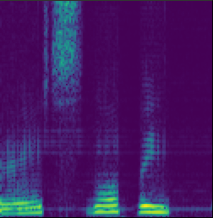}
            \caption[snr10]%
            {{\small $\hat{\bs}_1^{(6)}$ (SI-SDR: 13.29 dB)}}    
            \label{fig:snrrev}
        \end{subfigure}
        \hfill
        \centering
        \begin{subfigure}[t]{0.195\textwidth}
            \centering
            \includegraphics[height=1.220in]{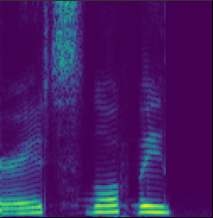}\vspace{0.03in}
            \caption[snr-5]%
            {{\small $\bs_1^{~}$}}    
            \label{fig:snr-5}
        \end{subfigure}
        
        \vskip\baselineskip
        
       \begin{subfigure}[t]{0.195\textwidth}
            \centering
            \includegraphics[height=1.220in]{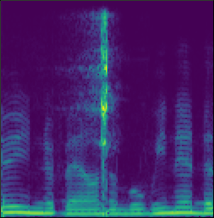}\vspace{0.035in}
            \caption[snr-5]%
            {{\small $\bx_2^{~}$ (SI-SDR: 3.05 dB)}}    
            \label{fig:snr-5}
        \end{subfigure}
        \hfill
        \begin{subfigure}[t]{0.195\textwidth}  
            \centering 
            \includegraphics[height=1.220in]{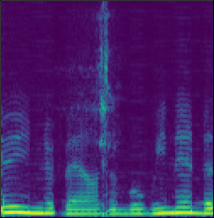}
            \caption[snr0]%
            {{\small $\hat{\bs}_2^{(2)}$ (SI-SDR: 14.22 dB)}}    
            \label{fig:snr0}
        \end{subfigure}
        \hfill
        \begin{subfigure}[t]{0.195\textwidth}   
            \centering 
            \includegraphics[height=1.220in]{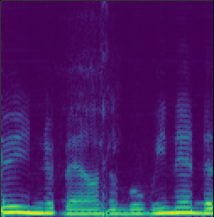}
            \caption[snr5]%
            {{\small $\hat{\bs}_2^{(4)}$ (SI-SDR: 19.93 dB)}}    
            \label{fig:snr5}
        \end{subfigure}
        \hfill
        \begin{subfigure}[t]{0.195\textwidth}   
            \centering 
            \includegraphics[height=1.220in]{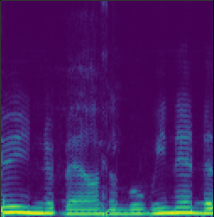}
            \caption[snr10]%
            {{\small $\hat{\bs}_2^{(6)}$ (SI-SDR: 19.98 dB)}}    
            \label{fig:snrrev}  
        \end{subfigure}
        \hfill
        \centering
        \begin{subfigure}[t]{0.195\textwidth}
            \centering
            \includegraphics[height=1.220in]{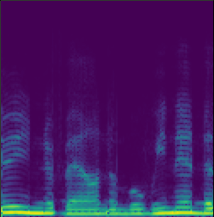}\vspace{0.035in}
            \caption[snr-5]%
            {{\small $\bs_2^{~}$}}    
            \label{fig:snr-5}
        \end{subfigure}
        \hfill
        \vskip\baselineskip
        \caption[something]
        {\small Denoising output samples from intermediate blocks. Each row represents a different example. The columns represent noisy mixture, estimated reconstructions from $l$-th blocks, and the corresponding ground-truth target clean speech.} 
        \label{fig:box}
    \end{figure*}


\section{Experimental Results and Discussion}
\label{sec:results}


Table \ref{tab:perf} presents the results of the competing systems.
We first compare BLOOM-Net with Baseline 1. 
At a glance, the model trained under \textit{Baseline 1 - Full} achieves the highest performance when it fully utilizes all 6 separator blocks. 
However, the model's scalability is limited, as it is evidently unable to perform denoising when less than 6 separator blocks are used (columns from $\ell=1$ to $5$). This behavior is expected since all 6 modules are trained altogether in the conventional end-to-end fashion---its intermediate separator outputs $\bz^{(l)}$ are not suitable for the shared masker module to compute the mask from, except for the final separator output $\bz^{(6)}$. 

Hence, it is unavoidable for Baseline 1 to train multiple versions of different block configurations to scale to various application and hardware requirements, which is the \textit{Baseline 1 - Int.} setup. Since in this set up there are totally six different end-to-end models, each of which specializes in each choice of $\ell$, they form the performance upper bound. However, the system's total spatial complexity is the sum of all six versions, which is not the most efficient option. 

BLOOM-Net, on the other hand, exhibits desired scalability. Instead of completely failing, \textit{BLOOM} shows decent performance at $\ell=5$ case, which is only a 0.1 dB drop from its $\ell=6$ setup. Its performance apparently drops as the model complexity decreases more. 
In addition, \textit{BLOOM}'s most powerful setup $\ell=6$ is suboptimal compared to Baseline 1. 

\textit{BLOOM-FT} addresses this issue by fine-tuning the entire modules via the sum of all blockwise loss functions. Compared against the oracle end-to-end Baseline 1 results (\textit{Baseline 1 - Int.)}, the fine-tuned results usually show less than 0.3 dB drop in all cases. Hence, we claim that the properly fine-tuned BLOOM-Net is almost comparable to the traditional end-to-end models while it provides unprecedented scalability and spatial efficiency.

Next, we draw attention towards the time-domain blockwise optimization method, \textit{Baseline 2}. 
Compared to BLOOM-Net that operates in the latent domain, \textit{Baseline 2} sequentially feeds time-domain inputs to the next stand-alone module. 
This incurs an overhead of learning the feature transformation and its inverse operation, constraining each block to only learn a shallow latent representation; thus, the improvement by adding more weak separation blocks is only minimal.
It showcases the merit of the proposed BLOOM-Net algorithm that performs blockwise optimization in the latent space. 

Fig. \ref{fig:box} shows two denoising examples where the different model complexity choices affect the quality of the output. 
The deeper the inputs are processed within the BLOOM-Net, the higher the denoising quality. 
The depth of the network can be decided based on the requirements of the deployed environment or users' preference. This demonstrates BLOOM-Net's scalability feature that offers flexible quality-complexity tradeoff to adjust to the test environment. 

Finally, Table \ref{tab:cmplxty} shows the inference-time computational complexity of \textit{Baseline 1 - Int.} and BLOOM-Net in terms of number of parameters and multiply-accumulate (MAC) operations. Note that \textit{BLOOM} and \textit{BLOOM-FT} are equivalent in this context. 
First, there is no difference in computational complexity (MACs), as the active modules during inference are the same. 
However, BLOOM-Net exhibits a significant advantage in terms of spatial complexity when we assume a scalable model. For example, if the device can afford up to $\ell=2$, the baseline has to prepare two different versions with $L=1$ and $L=2$ for the best performance in both energy-efficient and performance-boosted use cases. In doing so, although these two versions' sizes are $0.28M$ and $0.42M$ parameters, respectively, their sum amounts to $0.70M$. Likewise, in order for Baseline 1 to be scalable, it exponentially accumulates spatial complexity as $\ell$ grows. Conversely, BLOOM-Net manages this scalability issue more carefully: it increases the model size just by the amount of a single separator block and its insignificantly small block-specific masker and decoder modules (about 0.21M parameters). Hence, for example, our scalable BLOOM-Net with five separators (1.13M) is smaller than the baseline that covers three complexity profiles (1.26M).

\renewcommand{\arraystretch}{1.}
\begin{table}[t]
\centering
\caption{Computational requirements of time-domain ResNet models trained under Baseline 1 and our proposed BLOOM-Net method. The number of parameters reported encompasses the entire model parameters needed to implement the scalable model. 
MACs are computed given 1-second inputs.
}
\resizebox{\columnwidth}{!}{
\begin{tabular}{cccccccc}
\hline
& Method & $\ell=1$ & $\ell=2$ & $\ell=3$ & $\ell=4$ & $\ell=5$ & $\ell=6$\\
\hline\hline
\multirow{2}{*}{MACs (G)} & Baseline 1 & \multirow{2}{*}{0.53} & \multirow{2}{*}{0.80} & \multirow{2}{*}{1.07} & \multirow{2}{*}{1.33} & \multirow{2}{*}{1.60} & \multirow{2}{*}{1.87} \\
& BLOOM & & & & & & \\
\hline
\multirow{2}{*}{Params (M)} & Baseline 1 & 0.28 & 0.70 & 1.26 & 1.95 & 2.78 & 3.64\\
 & BLOOM & 0.28 & 0.49 & 0.71 & 0.92 & 1.13 & 1.34\\
\hline
\end{tabular}
\label{tab:cmplxty}
}
\end{table}

\section{Conclusion}
\label{sec:conc}
In this study, we introduced BLOOM-Net, a novel algorithm for scalable speech denoising. We postulated that scalable implementation of a deep learning-based speech enhancement system is critical to handle various resource-related test conditions that a device faces. While traditional end-to-end time-domain source separation models have shown advanced separation performance, we claimed that such a system cannot provide the desired scalability. Our BLOOM-Net is with a carefully designed residual learning scheme that performs blockwise optimization to improve the model performance in an incremental way. Since the blockwise optimization was on each individual separator module that greedily contributes to the model performance, BLOOM-Net achieves the scalablility---the device can freely choose from the multiple profiles based on its resource constraint. In doing so, the enhancement quality is reasonably associated with the model complexity. Source codes are available at \href{https://saige.sice.indiana.edu/research-projects/bloom-net}{https://saige.sice.indiana.edu/research-projects/bloom-net}.




\newpage

\bibliographystyle{IEEEtran}
\bibliography{mjkim}


\end{document}